# Urban-Rural Environmental Gradient in a Developing City: Testing ENVI GIS Functionality

*Lemenkova P.A.*

*(Charles University in Prague, Faculty of Science, Institute for Environmental Studies, Czech Republic)*

Abstract. The research performs urban ecosystem analysis supported by ENVI GIS by integrated studies on land cover types and geospatial modeling of Taipei city. The paper deals with the role of anthropogenic pressure on the structure of the landscape and change of land cover types. Methods included assessment of the impact from anthropogenic activities on the natural ecosystems, evaluation of the rate and scale of landscape dynamics using remote sensing data and GIS. The research aims to assist environmentalists and city planners to evaluate strategies for specific objectives of urban development in Taiwan, China.

The study aims at the assessment of the landscape development and environmental analysis of the Taipei city based on GIS and remote sensing methods. Methodology includes analysis of the environmental settings of the Taiwan region (land cover characteristics, climatic and topographic settings) and anthropogenic factors of the social landscapes. Research objectives included GIS based geo-visualization, spatial analysis of the raster data, i.e. satellite images and. Different factors involved in the landscape formation were studied and analyzed, both individually and in combination. The paper studies urban processes in Taipei in general and how GIS and remote sensing data can be used for the spatial analysis. The research deals with classification and analysis of the selected landscapes in Taiwanese ecosystems, and suggest development of the effective methods for wetlands assessment using a multi-metric approach. The research included analysis of the degree of landscapes disturbance according to the calculated metrics index values for specific land-use categories within the landscapes. As far as conservation problems are concerned, two scientific works place particular emphasis on the urban ecosystems on Taipei [9]. These are «Pattern and divergence of tree communities in Taipei's main urban green spaces» and «Diversity and distribution of landscape trees in the compact Asian city of Taipei», issued recently. The papers deal with the analysis of the landscape biogeographic characteristics of the urban forests: species diversity, composition and richness, spatial variability.

The interplay between both parts of the ecosystem, i.e. the anthropogenic impacts of the global city on the one side, and ecological patterns and processes on the other, is aimed to achieve balanced mutual co-existence within the global ecosystems. Modified land use types affect hydrological components in the surrounding watersheds and results in changes of the land use patterns, namely, landscape composition, configuration, connectivity, variety and abundance of patch types within a landscape, which are the primary descriptors of the landscape pattern. The complex functioning, structure and inter-relationships of the landscape components in urban ecosystem are largely influenced by the anthropogenic activities, inasmuch as the population concentrate in the metropolises and large cities. The location of Taipei in the southern subtropical humid climatic zone largely influences its environmental settings. Various disasters and hazards associated with specific climate settings such as torrential rains, occasional floods, rainstorms, seasonal typhoons have serious impacts and environmental consequences on the city functioning. Frequent and intense typhoons together with stormy rains and floods are natural disasters specific for Taiwan.

Consequences of hazards include significant property damage, destroyed farmlands, crashed roads, partially or completely destroyed houses, human injures, diseases and lives losses. Urban landscapes of Taipei are being formed during long historical process, as a result of the ongoing entangled interactions between natural forces and anthropogenic social factors.

However, the process of urbanization becomes more and more notable in the Taiwan in the past decades. Intense urbanization creates new conditions for the human-nature co-existence. It triggers environmental changes that occur recently. Examples include transformation of natural landscapes, changes in landscape diversity, deforestation, wetlands destruction, landscape fragmentation to name a few. Interactions between typhoons, urban sprawl and economic development significantly affect urban functioning in Taipei. The typical functions of urban ecosystem include cultural, supporting and regulating services which indicate core criteria for the measurement of the environmental quality of urban areas [7]. The complexity of these functions and suggested four aspects of the urban landscapes: spatial (architecture and planning space, organization and access to green spaces within the city), human (i.e. people and social interrelations), functional (recreational, commercial, transport services and welfare) and contextual aspects (life style and environmental health) [2]. Cultural aspect of the urban landscapes includes spaces for recreation, property and community value. Supporting landscape function involves biodiversity, green space and soil quality, energy flows, storage and cycling of nutrients and materials. To achieve harmonious co-existence between various components urban ecosystem should maintain environmentally appropriate conditions: cleanness, landscape aesthetics and attractiveness. The effects of urbanization on natural landscapes are studied in substantial existing literature [6], [19], [20], [14], [8], [16].

Methodologically, current study is focused on technical approach of raster data processing. There are numerous classification algorithms and techniques that determine natural spectral groups from the initial pixels sets. For instance, the most well-known are Parallelepiped classification, Neural Nets, Decision Trees, Mahalanobis Distance, Minimum Distance, and Maximum Likelihood classifiers, ISOCLUST, K-means (Figure 1). Usually, it is not easy to decide, which classifier method is *a priori* the best cartographic solution for actual research problem, due to different factors that vary significantly: characteristics of raster images, mapping scales, specific situation of the study area, reflectance properties of the local land cover types, landscape structure and heterogeneity.

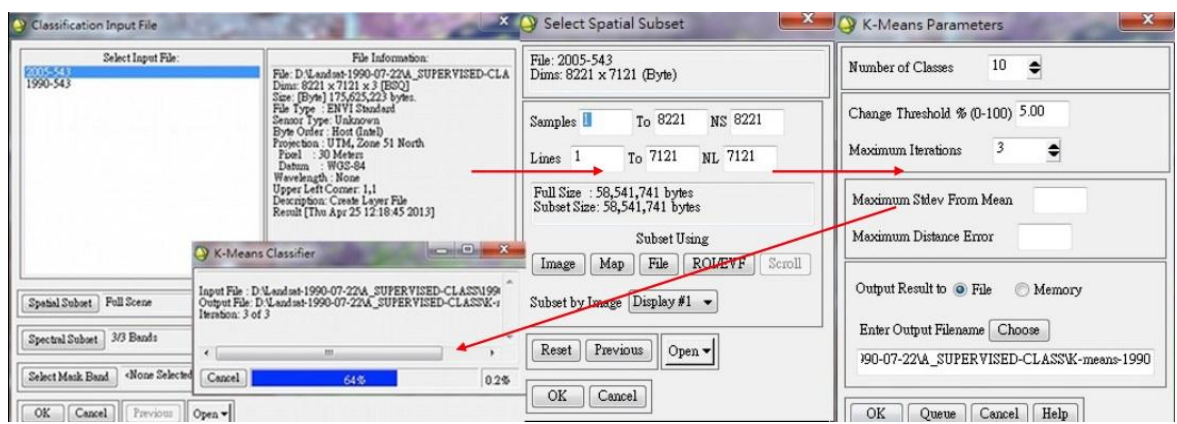

**Figure 1 – Classification process by K-Means classifier algorithm, ENVI**

A report on the statistics was generated and recorded as an 'Error Matrix', 'Kappa Statistics' and 'Accuracy'. Afterwards, to visually inspect classification process, the selected areas with the most diverse landscape structure and high heterogeneity of the land cover types were verified and validated by the overlapping of the Google Earth using activated function "connect to Google Earth". The "Link Google Earth to View" and "Sync Google Earth to View" options enabled to synchronize the view areas between the Google Earth and current image view. This enabled to check up highly heterogeneous areas where questions about land cover type arose. The land cover types were visually assessed and identified. The resulting clusters enabled to analyze spectral and textural characteristics of the land cover types in the selected area. Based on cluster segmentation, homogenous land cover classes were differentiated. The accuracy assessment was done using embedded statistical analysis in GIS, to estimate preciseness of the classification and preform post-classification (Figure 2).

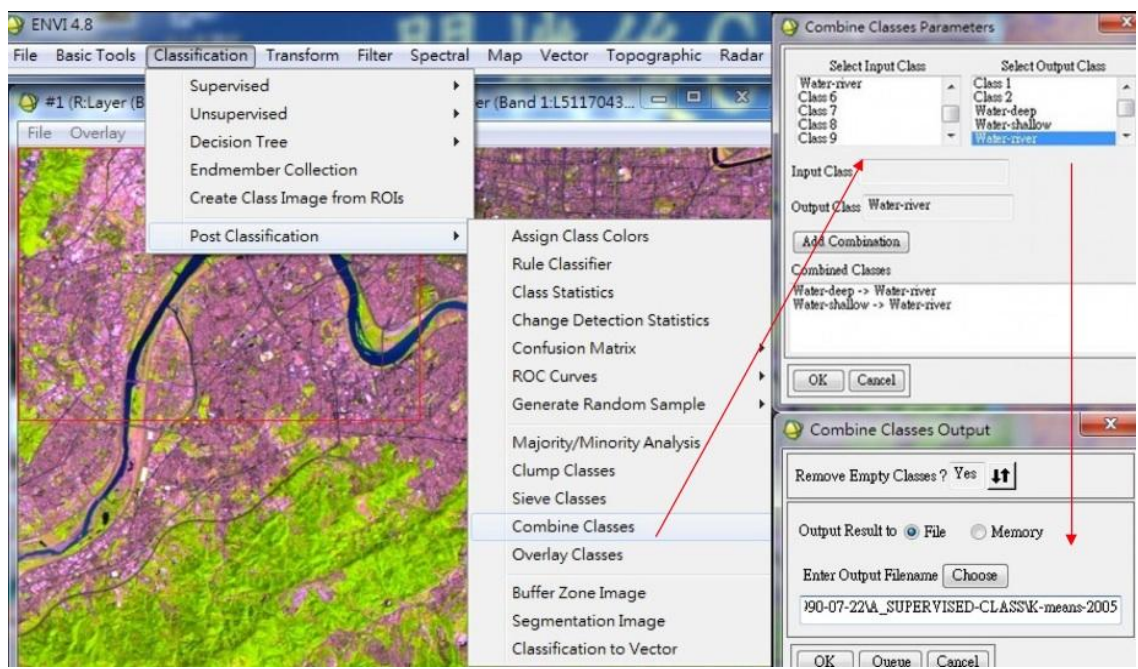

**Figure 2 – Post-classification editing: re-classifying and combining similar classes**

The study demonstrated successful application of GIS algorithms in combination with Landsat TM satellite imagery processing for identification of the land cover types in a rapidly developing areas where anthropogenic impacts have considerable effects. The results demonstrated (Table 1) most rapid increase of the urban spaces in the region 3 (from 24 % to 38%) which is explained by favorable location of the city outskirts from which the labour immigrants come to the city for jobs and settlements.

**Table 1 – Calculated areas of the selected land cover types in Taipei: 1990 vs 2005**

| Year, % | I: city areas | I: city parks | I: forest | I: grass | II: city areas | II: city parks | II: forest | II: grass | III: city areas | III: city parks | III: forest | III: grass |
|---|---|---|---|---|---|---|---|---|---|---|---|---|
| 1990 | 17,1 | 5,7 | 19,8 | 21,4 | 79,3 | 4,3 | 32,7 | 15,9 | 24,3 | 3,7 | 42,4 | 13,8 |
| 2005 | 21,9 | 4,1 | 18,6 | 18,3 | 83,2 | 5,2 | 31,8 | 14,7 | 38,2 | 4,3 | 41,8 | 12,1 |

Region 2 is a core industrial part of the city that is not developing rapidly due to the limited space available for new buildings. Region 3 is an agricultural part of the city where rural households are traditionally developed. Areas of forests and grasslands are decreasing in all the three parts of the city which proves the intensified and rapidly developing urbanization. The development of various land cover types within the Taipei's landscapes is affected, besides its topography, by such environmental impacts as climatic and hydrological processes which influence soils moisture and, consequently, vegetation, e.g. in riverine estuaries, wetland areas and typhoons. The most notable feature of the climatic settings is, besides high mean temperature with annual mean temperature above 10C, and a high humidity throughout the year with annual precipitation between 2000-2600 mm/year [12]. This is explained by the ample rains with annual rainfall total between 1200-1700 mm, often rainstorms and typhoons which represent a typical phenomena of the south subtropical climate [4]. The calculation of the landscape metrics and indices using classical methods of landscape ecology allowed to highlight spatial pattern optimization within a landscape, to characterize differences between the planned alternatives, to produce a land cover pattern that focuses on the specific aims of the optimal land use planning [11]. The Taipei's urban parks accommodate the highest biodiversity richness, while the built-up areas with space limitations the poorest one, as represented by popular native species. Landscape fragmentation leads to the disrupted or decreased connectivity between the singular patches within one landscape [1], [10]. Furthermore, it degrades basic functions of the life-supporting environment through the modification of the energy and material flows and nutrient cycling of the ecosystems [12]. At the same time, optimization and management of land-use structure and landscape pattern are primarily based on the influence of land cover patterns on the ecosystem functioning [21].

Many factors induce land use changes in the urban areas. To name a few, geomorphic settings of the area (altitude and slope steepness), topology, topographic location (distance from rivers, roads, built-up areas, urban areas), physical-geographic characteristics (soil drainage, erosion coefficient) and social factors (population density, occupation). Land cover changes caused by the extensive anthropogenic impacts in urban areas affect the complexity of the terrestrial landscapes and cause losses in the ecosystem, increase its environmental instability and decrease resilience towards external effects [20]. Twofold factors influence urban land cover changes [15]: 1) nature of the land use, i.e. human activities; 2) level of spatial accumulation, i.e. intensity and concentration of the human activities. Predictably, more urbanized areas have considerably higher level of concentration of the human-affected land use, while suburban regions have lower levels of accumulation [15]. Needless to say how much more environmentally vulnerable is the situation of the capitals, such as Taipei, comparing to the province towns. This is a specific concern for Taipei, as rapidly growing city and high urbanization rate affect biodiversity and cause ecosystem sustainability [17], [18].

The paper discussed current issues of the highly specific south China, the possibilities of sustainable development of the city placed in highly vulnerable climatic and environmental conditions with significant human pressure. It gives a detailed review of current ecological situation of Taipei and examines two sub-categories of the sustainable co-existence between man and nature in sub-tropics: social and human impacts on the landscapes and natural factors determining modern city formation and development. A theoretical issues that surround the environmental problems of the urban landscapes of Taiwan and current economic situation of Taiwan and its geographic specific problems are discussed. The research shows that urban system of Taipei have particular specific points of concern due to its complexity, heterogeneity and interplay between man and nature. The study adopted GIS methods of the existing works focused on geospatial analysis using satellite images with applied Landsat TM. Many studies used digital GIS tools to analyze urban landscape structures, to detect spatio-temporal changes in land cover types, to assess spatial patterns of urban landscapes and highlight relationships among various components of the ecosystems. However, despite long history of the GIS, the effective application of the remote sensing data such as Landsat with regards to Taiwan have so far received scant attention. For example, one may use multispectral images to compare land cover patterns and to assess different degrees of urbanization in the areas of cities Tokyo, Kyoto and Taipei [25]. The landscape structure and vulnerability towards external environmental and human impacts is discussed which includes consequences of the uncontrolled urbanization for the environment. Paper applied GIS spatial analysis for the study area and how GIS can be used for studies of the spatio-temporal land cover changes. The spatiotemporal pattern of the landscape diversity in Taipei changed between 1971 and 2005. The main issue of the environmental impacts is changed landscape city patterns caused by modified land use during urbanization. Modified land use types affect various components of the ecosystem. Destroyed urban landscapes cause gradual destruction of the city as urban well-being and functioning strongly depends on the natural resources: availability and quality of the diverse energy and materials. Though economic growth is crucial to the successful development of the country and well-being, the uncontrolled use of the natural resources and non-renewable energy cause unsustainable development of the complex urban ecosystems and can result in environmental consequences [3].